%% file: GammaBackgroundPaper.tex
\documentclass[twocolumn,epjc3]{svjour3}

\RequirePackage{mathptmx}      % use Times fonts if available on your TeX system
\RequirePackage{flushend}
\usepackage{graphicx}
\usepackage{hyperref}
\usepackage{pifont}
\usepackage{color}
\RequirePackage{booktabs} %Makes tables better?

\usepackage{cite}
\usepackage{amsmath,amssymb}
\usepackage{siunitx}
\usepackage{multirow}
\def\equationautorefname~#1\null{Eq.~(#1)\null}
\def\figurename{\bf{Fig.}}
\def\tablename{\bf{Table}}

\usepackage{upgreek}

\usepackage{lineno}

\DeclareSIUnit\cps{cps}
\DeclareSIUnit\gauss{G}
\DeclareSIUnit\CL{C.L.}
\journalname{Journal}

\begin{document}

%\linenumbers

\title{Gamma-induced background in the KATRIN main spectrometer}

\input{authors_epjc}

\date{\today}

\maketitle
%\nopagebreak{}
%\newpage

\begin{abstract}
The KATRIN experiment aims to measure the effective electron antineutrino mass $m_{\overline{\nu}_e}$ with a sensitivity of \SI{0.2}{\eV/c^2} using a gaseous tritium source combined with the MAC-E filter technique.
A low background rate is crucial to achieving the proposed sensitivity, and dedicated measurements have been performed to study possible sources of background electrons.
In this work, we test the hypothesis that gamma radiation from external radioactive sources significantly increases the rate of background events created in the main spectrometer (MS) and observed in the focal-plane detector. 
Using detailed simulations of the gamma flux in the experimental hall, combined with a series of experimental tests that artificially increased or decreased the local gamma flux to the MS, we set an upper limit of \SI{0.006}{count/\s}~(90\% C.L.) from this mechanism.
Our results indicate the effectiveness of the electrostatic and magnetic shielding used to block secondary electrons emitted from the inner surface of the MS.

\end{abstract}
%\nopagebreak{}

%\tableofcontents

\section{Introduction}
\label{sec:intro}

The Karlsruhe Tritium Neutrino (KATRIN) experiment has been designed to reach a sensitivity of \SI{0.2}{\eV/c^2} (90\% C.L.) on the effective electron antineutrino mass $m_{\overline{\nu}_e}$~\cite{KATRIN2005}.
As the successor to the Mainz~\cite{Kraus2005} and Troitsk~\cite{Aseev2011} experiments, KATRIN will precisely measure the energy of electrons produced from tritium $\upbeta$-decay and determine $m_{\overline{\nu}_e}$ by fitting the shape of the $\upbeta$-spectrum near the endpoint energy (\SI{18.6}{\keV}).
Reaching the sensitivity goal requires a thorough understanding and mitigation of background sources along the entire beamline of the experiment.
Background electrons produced in the main spectrometer (MS), the high-resolution MAC-E filter~\cite{Beamson1980, Lobashev1985, Picard1992} that analyzes the energy of the $\upbeta$-particles, are of particular importance. 
Several sources of background in the MS have already been analyzed in detail, including cosmic-ray muons~\cite{Altenmueller2019} and radon decays~\cite{RadonPaper}. 

X-rays and gammas can also contribute to the background and have been previously studied with other MAC-E filter spectrometers. 
The Mainz spectrometer~\cite{PhDFlatt2004} and the KATRIN pre-spectrometer~\cite{DthLammers2009} were both irradiated using an X-ray tube with a peak energy of $E_{\text{X-ray}}=\SI{70}{\keV}$.  
These tests showed the effectiveness of the electrostatic and magnetic shielding measures in place, which suppressed the background contribution caused by the X-ray tube by up to a factor of 50. 
However, the total background was still strongly elevated while the irradiation took place, indicating that gamma radiation can potentially have a large contribution to the background rate for MAC-E filter spectrometers.
In this paper, the effect of environmental gamma radiation on the MS background rate is investigated through a combination of simulation and measurement data, and the effectiveness of the shielding in the MS is demonstrated.

The experimental setup of the KATRIN experiment is described in Section~\ref{sec:apparatus}, focusing on the MS and the focal-plane detector (FPD) system, which measures the electron rate at the exit side of the MS.
Section~\ref{sec:EnvironmentalGamma} discusses the background generation due to environmental gammas and gives details about their simulation in the spectrometer hall.
In Section~\ref{sec:BackgroundMeasurements}, the simulation results are compared to background measurements under conditions of gamma enhancement and suppression.  
The contribution of environmental gamma radiation to the KATRIN background rate is derived in Section~\ref{sec:UpperLimit}, and some final remarks on the MS background are provided in Section~\ref{sec:conclusion}.

\section{Experimental apparatus}
\label{sec:apparatus}

\begin{figure*}
	\centering
	\vspace{10pt}
        \includegraphics[width=\textwidth]{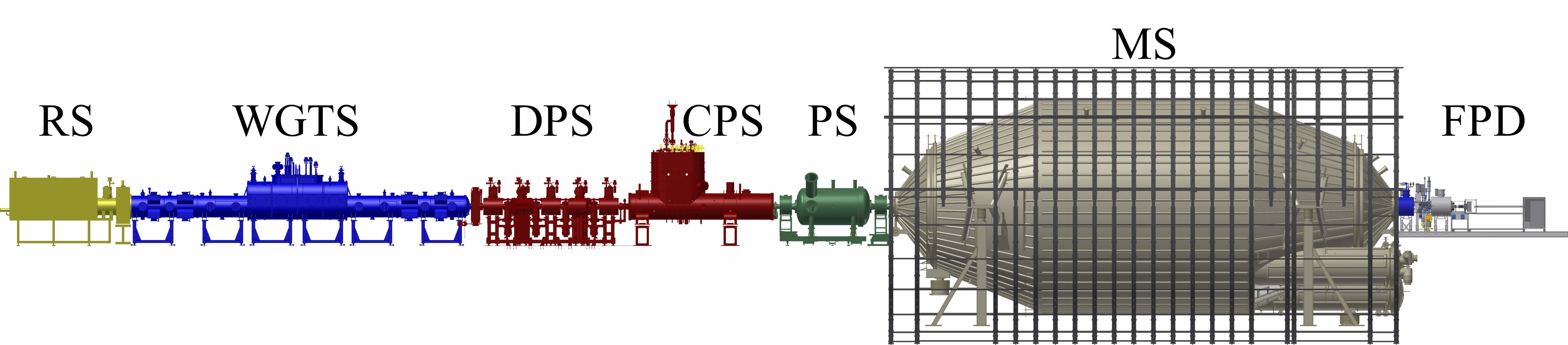}
        \caption{The KATRIN beamline.  
        From left to right: the rear section (RS), the windowless gaseous tritium source (WGTS), the differential pumping section (DPS), the cryogenic pumping section (CPS), the pre-spectrometer (PS), the main spectrometer (MS) with surrounding air-coils, and the focal-plane detector (FPD) system.}
        \label{fig:katrinBeamline}
\end{figure*}

The KATRIN experiment is located at the Karlsruhe Institute of Technology (KIT), Campus North, near Karlsruhe, Germany.
The approximately \num{70}-\si{\m} long beamline is shown in Fig.~\ref{fig:katrinBeamline}.
The windowless gaseous tritium source has been designed to produce more than \num{e11} $\upbeta$-particles per second~\cite{Babutzka2012}.
Superconducting magnets adiabatically guide emitted electrons through the differential~\cite{PhDKosmider2012} and cryogenic~\cite{Gil2010, PhDJansen2015} pumping sections, where the tritium flow must be reduced by 14~orders of magnitude~\cite{KATRIN2005}.

Two MAC-E filters are available to analyze the electron energy. 
The smaller pre-spectrometer can be used as a pre-filter to reflect low-energy $\upbeta$-particles back toward the source and only allow the highest energy electrons, close to the endpoint energy, to enter the MS.
The precise energy filtering is performed in the MS, which operates with an energy resolution of about \SI{1}{\eV} near the tritium endpoint~\cite{KATRIN2005}.
Electrons that pass through the MS are measured with the FPD system.
Due to their importance in studying the gamma-induced background, the MS and FPD system will be discussed in detail in the following subsections.

\subsection{Main spectrometer}
\label{subsec:ms}

The largest component of the KATRIN beamline is the MS (see Fig.~\ref{fig:katrinBeamline}).  
The stainless steel vessel has an inner diameter of \SI{9.8}{\m} at its center and a total length of \SI{23.3}{\m}~\cite{KATRIN2005}.
The walls of the MS vary in thickness between \SI{32}{\mm} for the central cylindrical region and \SI{25}{\mm} for the conical sections~\cite{KATRIN2005}.
During standard KATRIN operation, the retarding voltage $U_{\textrm{0}}$ applied to the MS hull will be varied systematically around \SI{-18.6}{\kV} in order to analyze electrons with energies close to the tritium endpoint energy.
To prevent scattering with residual gas, the MS is designed to operate under ultra-high vacuum conditions, with a pressure close to \SI{e-11}{\milli\bar}~\cite{Arenz2016}.

Superconducting solenoids placed at the entrance and exit of the MS generate the magnetic flux tube that is responsible for adiabatically guiding electrons through the vessel~\cite{Arenz2018a}.
Air-coils surrounding the MS are used to fine-tune the magnetic field inside the vessel and compensate for the Earth's magnetic field~\cite{Glueck2013,Erhard2017}.
By manipulating the magnitude and polarity of the air-coil currents, it is possible to create non-standard magnetic field settings inside the vessel (see Section~\ref{sec:BackgroundMeasurements}).

The magnetic field inside the MS acts as a passive shield against charged particles emitted from the vessel surface; the particles are deflected by the Lorentz force back toward the walls of the vessel, or they follow magnetic field lines that do not reach the detector.
Additional shielding is provided by a wire electrode system installed at the inner surface of the MS~\cite{Valerius2010}.
The inner electrodes (IE) can be placed at an offset potential $\Delta U_{\textrm{IE}}$ relative to the vessel (i.e. $U_{\textrm{IE}} = U_0 + \Updelta U_{\textrm{IE}}$).
If $\Updelta U_{\textrm{IE}}$ is set to a negative value, most low-energy electrons emitted from the vessel walls will be repelled by the wires back toward the surface.

\subsection{Focal-plane detector}
\label{subsec:fpd}

The FPD~\cite{Amsbaugh2015} is a monolithic, segmented silicon PIN diode that detects particles emerging from the MS. These particles pass through an ultra-high vacuum system mated to the exit of the MS. Charged particles can be further accelerated using a post-acceleration electrode (PAE), with voltage $U_{\textrm{PAE}}$, immediately preceding the detector wafer; this acceleration helps distinguish signal electrons from background originating within the FPD system. A superconducting solenoid focuses charged particles onto the wafer, which has a sensitive area with a diameter of \SI{90}{\mm}~\cite{Amsbaugh2015}. 
The dartboard-segmentation pattern of the wafer into 148 equal-area pixels provides sensitivity to the transverse spatial distribution of the particles.
The wafer is divided into 12 concentric rings, each with 12 pixels, and a central bullseye with four pixels.
The PIN diode is biased with a voltage $U_{\textrm{bias}}$; signals pass through preamplifiers (located in vacuum) before readout.

Energy and timing for each event are analyzed online in the data-acquisition system using a cascaded pair of trapezoidal filters. Energy calibration is provided by gammas from an $^{241}$Am calibration source. The electron response is characterized using a photo-electron source with adjustable energy~\cite{Amsbaugh2015}. Each source can be inserted into the line of sight of the FPD for a periodic, dedicated calibration run. An energy resolution of \SI{1.52}{\keV} (full width at half-maximum, FWHM) and timing resolution of \SI{246}{\ns} (FWHM) have been achieved with the system using \SI{18.6}{\keV} electrons and a \SI{6.4}{\micro\s} shaping time~\cite{Amsbaugh2015}.

\section{Environmental gamma radiation in the spectrometer hall}
\label{sec:EnvironmentalGamma}

During standard KATRIN operation, $\upbeta$-particles that are detected by the FPD must overcome the retarding potential applied to the MS.
This potential reaches its largest value near the middle of the vessel, approximately equidistant from both ends of the MS.
Because the value of the retarding potential will be scanned close to the endpoint energy, $\upbeta$-particles traveling through the middle region of the MS will have low kinetic energies, below about \SI{30}{\eV}~\cite{KATRIN2005}.
Because of the relatively poor energy resolution of the FPD compared to the MS, any low-energy secondary electrons produced in this region cannot be energetically distinguished from the signal $\upbeta$-particles.

When passing through the MS steel, environmental gamma radiation (i.e., primary radiation) can cause the emission of secondary particles, including secondary electrons.
Of particular concern is the generation of ``true-secondary'' electrons, which are defined to have energies below \SI{50}{\eV}~\cite{Furman2002}.
The inner surface of the MS provides a large area for electron emission: \SI{690}{\m^2} for the steel hull and \SI{532}{\m^2} for the wire electrode system (although the effective surface area for emission from the latter is reduced due to the two-layer structure of the IE)~\cite{Arenz2016}.
Due to imperfections in the magnetic and electrostatic shielding, secondary electrons emitted from these surfaces may have a small probability to enter the sensitive magnetic flux tube that connects to the FPD.
%As discussed in Section~\ref{sec:intro}, 
The validity of this background-generating mechanism has been confirmed with other MAC-E-filter spectrometers~\cite{PhDFlatt2004,DthLammers2009}.
Because of the increased size of the spectrometer, there is the potential for a significant background contribution from gamma-induced surface electrons in the MS.

\subsection{Radioactivity in the spectrometer hall}
\label{sec:siminput_radioactivity}

The MS was constructed using low-radioactivity materials in order to limit background production from environmental gammas.
The central cylindrical portion of the MS vessel was built from \SI{32}{\mm}-thick sheet metal, composed of type 316LN stainless steel.
A sample of this metal, from the same batch used for the MS vessel, was measured for radioisotopes; the results of this study are given in Table~\ref{table:activities}.
However, the primary source of gammas in the KATRIN spectrometer hall is not the vessel itself but rather the intrinsic radioactivity of the concrete used to construct the walls and floors.

The walls of the spectrometer hall are made of standard concrete.
To reduce the effect of ambient gamma radiation, the floors were built using a low-activity concrete due to their close proximity to the MS.
During construction of the hall, samples from every load of low-activity concrete were monitored with a NaI detector to ensure the activity fell within specifications.
Additionally, the gamma spectra of several samples of the concrete were measured at KIT using a shielded HPGe detector.  
The concrete activities shown in Table~\ref{table:activities} were derived from these measurements.  
$^{40}$K was found to be the largest contributor to the activity in the concrete, followed by the $^{238}$U and $^{232}$Th decay chains.

\begin{table}
 \caption{Specific activities (in units of \si{\becquerel/\kg}) determined from radioassay measurements for materials in the KATRIN spectrometer hall. 
 The activity of the wall concrete was only measured for the basement walls; for the upper walls, the same distribution of activities was assumed.
      The steel activities were measured at the Oroville Low Background Facility~\cite{Smith2015}.}
     \label{table:activities}
    \begin{tabular*}{\columnwidth}{@{\extracolsep{\fill}}lccc@{}}
    \hline
 Isotope & Wall concrete & Floor concrete & MS steel\\ 
\hline
$^{40}$K & \num{409 \pm 22} & \num{61 \pm 3} & $<(\num{3e-6})$ \\
$^{238}$U & \num{23 \pm 5}  & \num{6.0 \pm 0.7} & \num{5.0 \pm 2.5 e-2} \\
$^{232}$Th & \num{18 \pm 1} & \num{3.3 \pm 0.3} & $<(\num{1.2e-3})$ \\
$^{235}$U & \num{1.7 \pm 0.5}  & \num{0.6 \pm 0.1} & \\
$^{60}$Co &  &  & \num{1.8 \pm 0.3 e-3}\\
\hline
    \end{tabular*}
\end{table}

\subsection{Simulation of the gamma flux}
\label{sec:simdetails}

\begin{figure}
	\centering
         \includegraphics[width=\columnwidth]{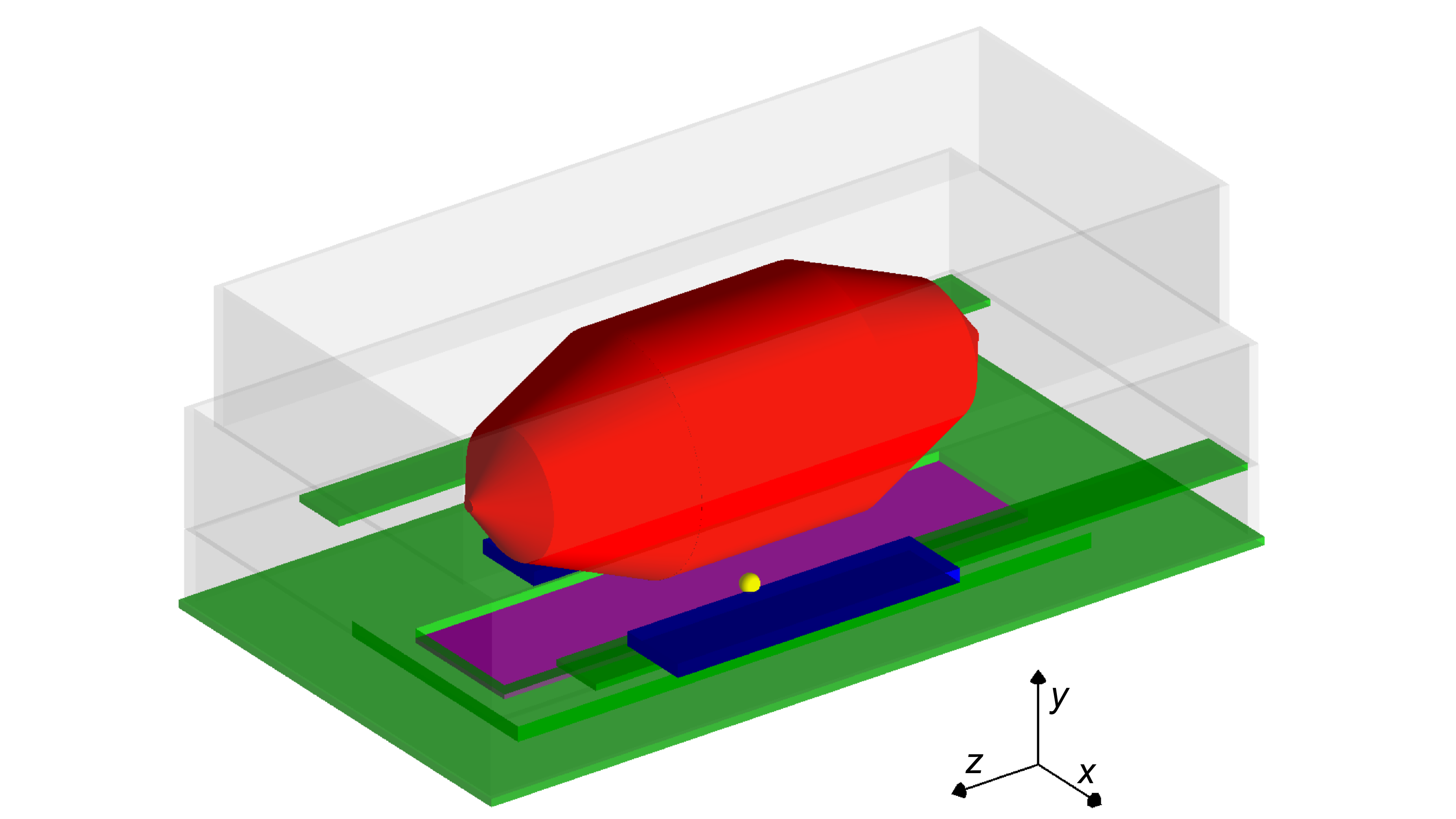}
        \caption{The \textsc{Geant4} simulation geometry, consisting of the steel MS (red), the concrete walls (gray) and floors (green), and, optionally, water tanks (blue) and basin water (purple).
        The yellow sphere indicates the position of the optional $^{60}$Co source.
        %The orange cube (not used in the simulation) shows the position of the HPGe detector when collecting the measured spectrum shown in Fig.~\ref{fig:energySpectrum}.
        %The orange disk shows the position of the HPGe detector when collecting the measured spectrum shown in Fig.~\ref{fig:energySpectrum} and was used to compute the flux described in the text \comment{to do}.
        The coordinate axes for the experiment are also shown, where the $z$-direction ($x$-direction) corresponds to due north (west).
        For orientation, the FPD is located north of the spectrometer.}
        \label{fig:geant4Geometry}
\end{figure}

To better understand the background due to environmental gamma radiation, simulations of the gamma flux in the MS were performed with the \textsc{Geant4} simulation toolkit~\cite{geant:2003, geant:2006, geant:2016}, version 10.4.p02.
A simplified reproduction of the KATRIN spectrometer hall was implemented; the geometry included the steel MS vessel and the concrete walls and floors (Fig.~\ref{fig:geant4Geometry}).  
Radioactive isotopes were uniformly created in the walls, floors, and steel with the relative production rates determined from the radioassay measurements given in Table~\ref{table:activities}. 
The effect of $^{222}$Rn in the air was also included, assuming an activity of \SI[multi-part-units = single]{49 \pm 15}{\becquerel/\m^3}, which is the average indoor radon level in Germany~\cite{WHO2007}.
The decays of the isotopes and any subsequent daughters were handled by the \textsc{Geant4} Radioactive Decay Module, and secular equilibrium was assumed.
This module has recently been updated by the \textsc{Geant4} collaboration to include newer versions of the Evaluated Nuclear Structure Data File (ENSDF) datasets~\cite{ENSDF} and to better ensure energy conservation for decays~\cite{geant:2016}.
Available physics processes were set by the \texttt{Shielding\_EMZ} physics list, which uses the most accurate electromagnetic physics models and is well-suited for shielding simulations~\cite{geant:PhysicsList}.

\begin{table}
      \caption{Simulated total gamma fluxes integrated at the inner surface of the MS. 
      For each component, the flux was calculated based on the activities listed in \autoref{table:activities}, after summing the individual contributions from the different radioisotopes.
      The errors are purely systematic and indicate the uncertainty of the component activities.}
     \label{table:SimulatedGammaFluxes}
\begin{tabular*}{\columnwidth}{@{\extracolsep{\fill}}lcr@{}}
   \hline 
Component & MS flux (gammas/s) & Percentage \\ 
\hline 
Concrete walls & \num{2.06 \pm 0.08 e6} & 80.2 \% \\
Concrete floors & \num{4.1 \pm 0.2 e5} & 15.8 \% \\
Air & \num{9.0 \pm 2.7 e4} & 3.5 \% \\
Steel hull & \num{1.1 \pm 0.5 e4} & 0.4 \% \\
\hline
Total & \num{2.56 \pm 0.09 e6} & \\
\hline 
    \end{tabular*}
\end{table}

The gamma fluxes determined from simulation are listed in Table~\ref{table:SimulatedGammaFluxes} for each geometrical component.
As expected, the concrete walls are the primary contributor to the gamma flux inside the MS, followed by the concrete floors.
The gammas originating from the steel vessel and the air, combined together, make up less than \SI{4}{\%} of the flux inside the MS.
%\comment{This paragraph to be updated.}
%To validate the simulation, a couple of comparisons were made to measurement data.
%After construction of the spectrometer hall, the environmental gamma radiation was measured using a NaI detector.
%A rate of \SI{5200 \pm 1500}{photons\per\square\m\per\s} was observed in the hall, for photons with energies between \SI{1}{\keV} and \SI{3000}{\keV}~\cite{PhDLeiber2014}.
%From the \textsc{Geant4} simulation, a rate of \SI{5090 \pm 22}{photons\per\square\m\per\s} was determined, which matches the measured value.

\begin{figure}
	\centering
        \includegraphics[width=\columnwidth]{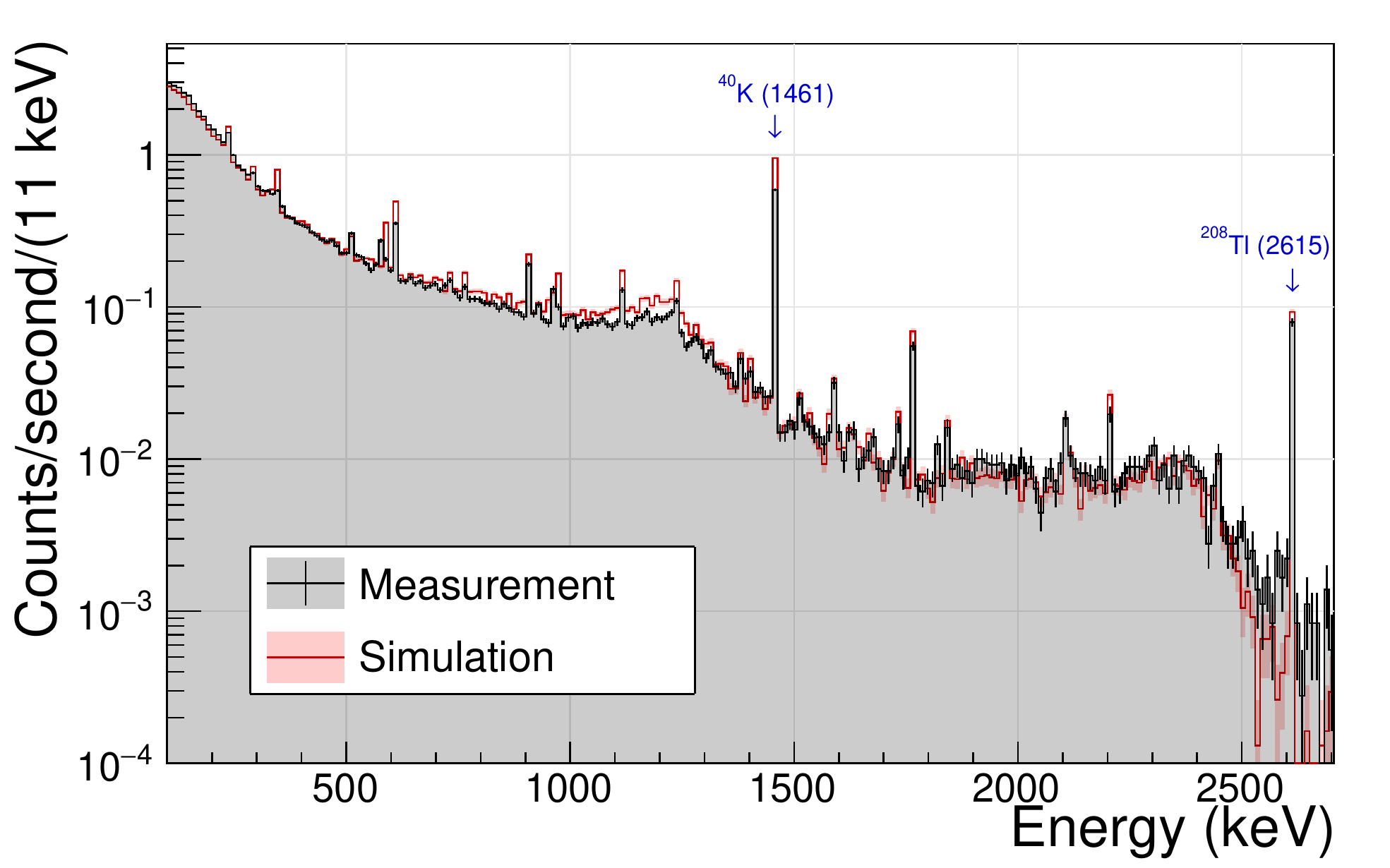}
        \caption{The energy spectrum measured by a HPGe detector in the spectrometer hall (grey), and the simulated spectrum for energy deposited in a germanium crystal by gammas originating from a \SI{0.6}{\m} by \SI{0.6}{\m} region of the concrete wall (red).
        The simulated spectrum has been normalized to the measured spectrum, with a normalization factor of \num{5.8}.
        Prominent lines from $^{208}$Tl ($^{232}$Th decay chain) and $^{40}$K are labeled for reference.}
        \label{fig:energySpectrum}
\end{figure}

%Later, 
Gamma spectra were measured at various locations in the spectrometer hall using a HPGe detector.
One of these spectra, collected for the detector facing the western wall in the basement of the hall, is shown in Fig.~\ref{fig:energySpectrum}. 
The dominant contributor to the gamma spectrum is the \SI{1461}{\kilo\eV} line from the decay of $^{40}$K~\cite{Chen2017}.
To compare with the measurement, a very simple germanium detector was implemented in the \textsc{Geant4} simulation.
For reasons of computational expense, only gammas from a small ($\SI{0.6}{m}\times\SI{0.6}{m}$) region of the concrete wall were simulated, and detector efficiencies were not accounted for in the simulation.
A scaling factor was therefore applied to match the measured rate.
Nonetheless, the simulated spectrum (see Fig.~\ref{fig:energySpectrum}) qualitatively replicates the important features of the measured gamma spectrum.
%, even though the latter suffers from detector effects~\cite{Gilmore2008Chap2} which were not accounted for in the simulation.
%Overall, the simulation is able to adequately approximate both the rate and spectrum of environmental gammas in the spectrometer hall.
Overall, the simulation is able to adequately approximate the spectrum of environmental gammas in the spectrometer hall.

\section{Background measurements}
\label{sec:BackgroundMeasurements}

During the summer of 2015, KATRIN proceeded with a measurement phase with only the MS and the FPD system, lasting several months.
One of the primary goals of this commissioning campaign was to study background events originating from the MS.
%Prior to the start of the measurements, the MS vessel was baked at \SI{200}{\degreeCelsius} for over a week to reduce the outgassing from the interior surface~\cite{PhDHarms2015}.

As described in the following subsections, the electron rate was measured by the FPD for several configurations that modified the flux of environmental gammas at the MS.
The voltage settings used for these measurements are given in Table~\ref{table:VoltageSettings}. 
%Because of the finite energy resolution of the detector system, the determination of an electron rate requires using an electron energy region of interest (ROI).  
%The default ROI is defined to be the energy range between \SI{3}{\kilo\eV} below and \SI{2}{\kilo\eV} above the expected electron energy at the detector.  
The electron rates were measured for the energy region of interest (ROI), which was defined as the range between \SI{3}{\kilo\eV} below and \SI{2}{\kilo\eV} above the expected electron energy at the detector.  
An asymmetric energy window is applied since the shape of the electron peak has a low-energy tail.
For an electron with initial energy $E_{\textrm{0}}\approx\SI{0}{\eV}$ emitted at the MS surface, the expected energy is set by the change in the electric potential between the detector wafer and the MS inner electrode: 
\begin{equation}
%\begin{aligned}
%E_{\textrm{expected}}&=E_{\textrm{0}}+e[U_{\textrm{PAE}}+U_{\textrm{bias}}-U_{\textrm{0}}-\Updelta U_{\textrm{IE}}],\\
E_{\textrm{expected}}=E_{\textrm{0}}+e[U_{\textrm{PAE}}+U_{\textrm{bias}}-U_{\textrm{0}}-\Updelta U_{\textrm{IE}}],\\
%&\approx\SI{28.7}{\kilo\eV}.
%\end{aligned}
 \label{eq:ExpectedEnergy}
\end{equation}
From the values given in Table~\ref{table:VoltageSettings}, the expected electron energy is \SI{\sim28.7}{\kilo\eV}.
%All electron rates listed in the text were determined using the default electron ROI, assuming $E_{\textrm{0}}\approx\SI{0}{\eV}$.  
Due to a broken preamplifier module in the detector readout, six FPD pixels (each located in a different pixel ring) were not functional.
The rate for each detector pixel ring with a missing pixel is linearly scaled by a corrective factor of $\frac{12}{11}$.

\begin{table}
      \caption{The voltage settings used during the gamma-induced background measurements.}
     \label{table:VoltageSettings}
    \begin{tabular*}{\columnwidth}{@{\extracolsep{\fill}}llr@{}}
    \hline
Component & Voltage & Value/range (\si{\V}) \\ \hline
MS hull & $U_{\textrm{0}}$ & \numrange{-18500}{-18600} \\
MS inner electrode & $\Updelta U_{\textrm{IE}}$ & \numrange{0}{-100} \\
Post-acceleration electrode & $U_{\textrm{PAE}}$ & \num{10000} \\
Bias ring of detector wafer & $U_{\textrm{bias}}$ & \num{120}  \\
\hline
    \end{tabular*}
\end{table}

\begin{figure}
  \centering
    \includegraphics[width=\columnwidth]{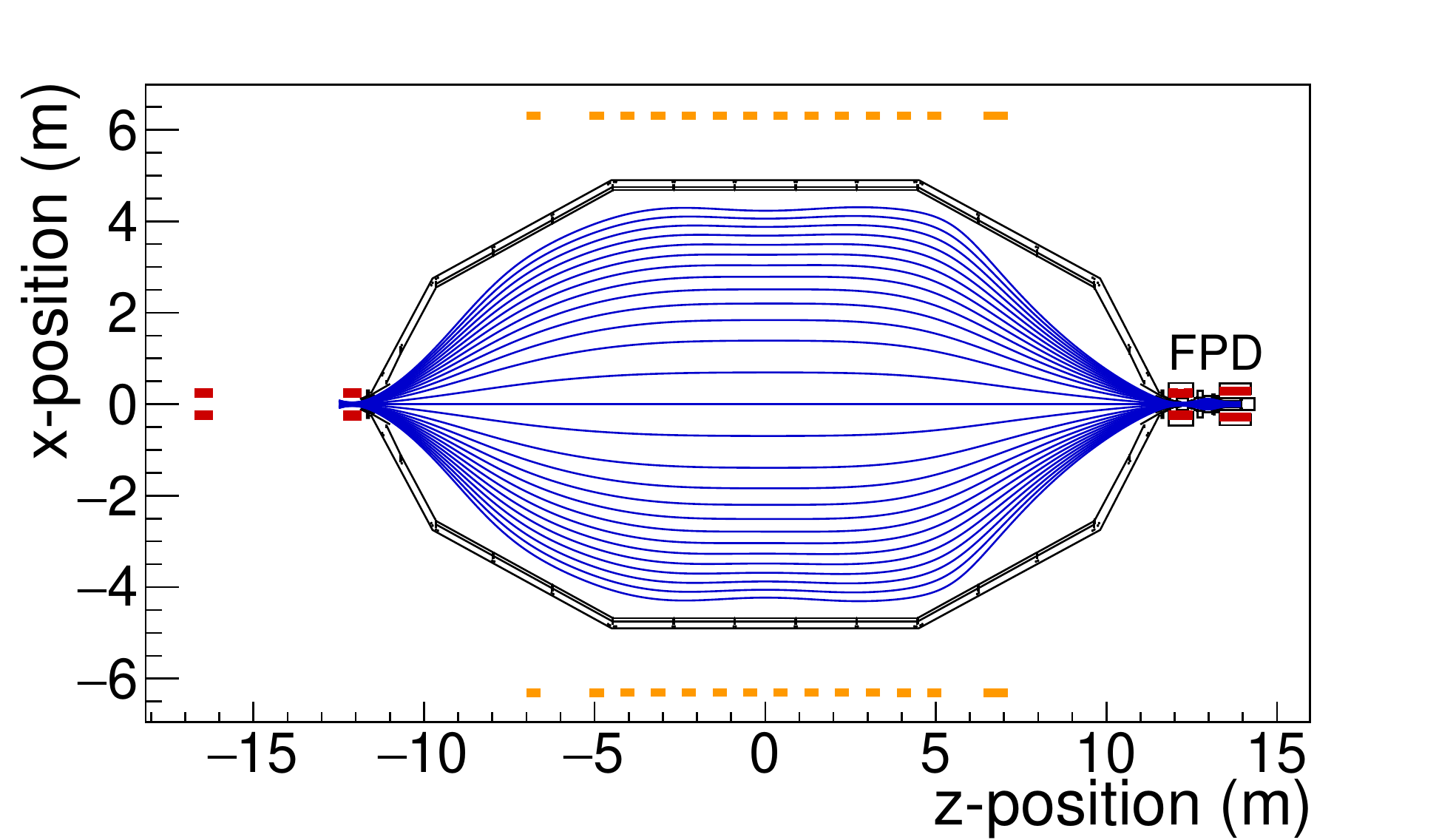}
    \includegraphics[width=\columnwidth]{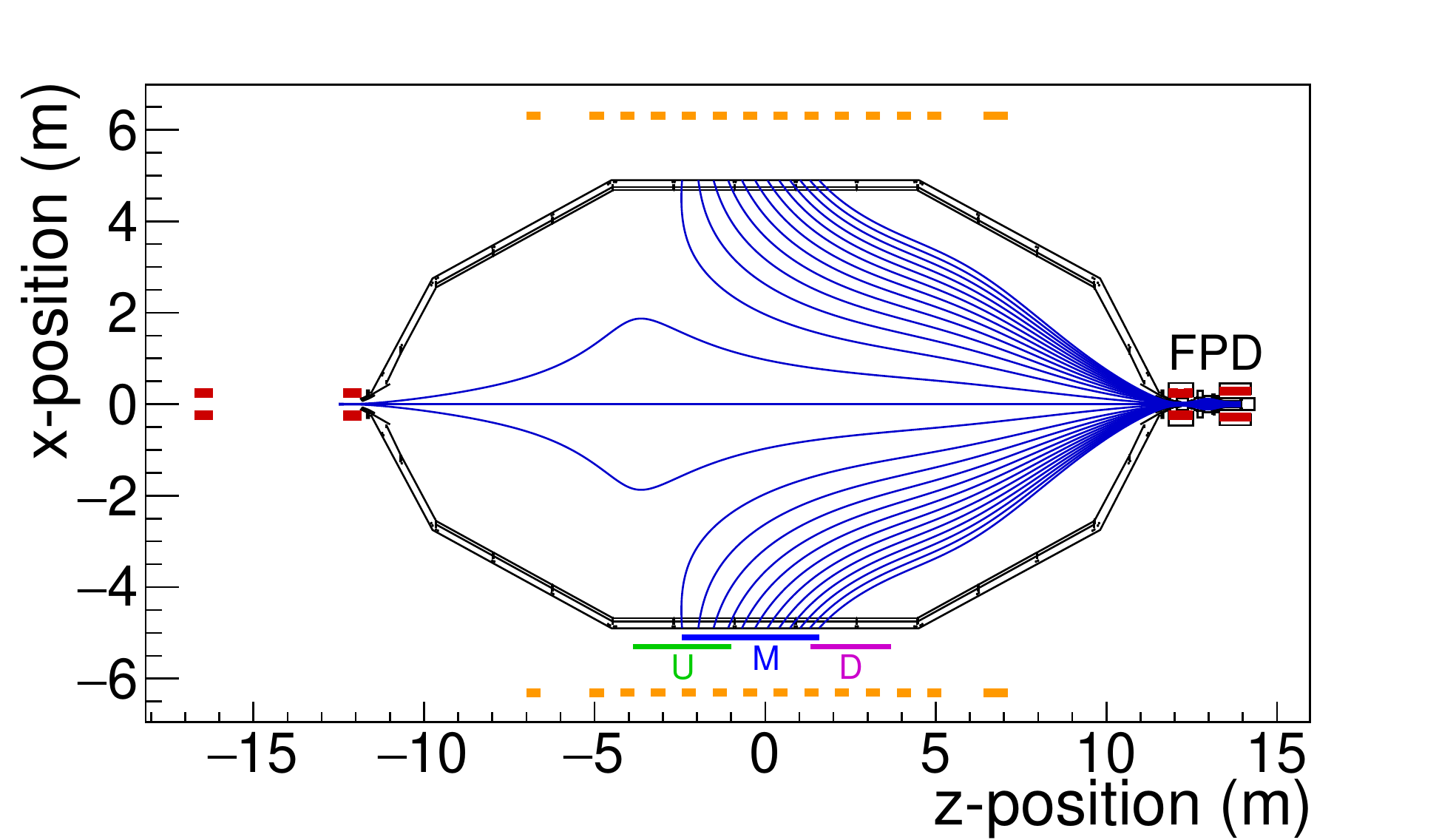}
      \caption{Symmetric (upper panel) and asymmetric (lower panel) magnetic field configurations inside the MS.  
      The displayed magnetic field lines (blue) intersect the FPD pixel rings, and are produced by the beamline solenoids (red) and air-coils (orange).
      This asymmetric configuration is given the name ``Asym. M'' since the field lines intersect the middle area of the MS ($\SI{-2.4}{\m}<z<\SI{1.6}{\m}$).
      This z-range (as well as the ranges for Asym. U and Asym. D, whose magnetic field lines are not drawn) is shown by the labeled, colored line near the bottom of the figure.
      }
    \label{fig:MagneticFieldSettings}
     \end{figure}

Several magnetic field configurations were utilized when studying the gamma-induced background.  
A symmetric magnetic field, similar to the planned field for standard KATRIN operation, is shown in the upper panel of Fig.~\ref{fig:MagneticFieldSettings}.
In this setting, the vast majority of electrons emitted from the MS surface should not reach the detector due to the inherent magnetic shielding of the flux tube~\cite{PhDWandkowsky2013}.
With this symmetric field setting, a total background rate of $0.561 \pm 0.005$~counts per second (cps) was measured.

In order to directly study electrons emitted from the surface, an asymmetric field configuration can be used by adjusting the currents applied to the air-coils surrounding the MS; an example field is shown in the lower panel of Fig.~\ref{fig:MagneticFieldSettings}.
Because the magnetic field lines directly connect the vessel walls to the FPD, this setting allows a significantly larger fraction of surface electrons to be detected. 
To ensure the measured electrons are emitted from a well-defined surface region of the MS, certain detector pixels are excluded from the rate analysis.
For the Asym. M configuration, the inner 16 detector pixels are excluded, while the inner 28 (4) pixels are excluded for the Asym. U (Asym. D) configuration.

\subsection{Enhancement of gamma flux}
\label{subsec:Co60-enhancement}

\begin{figure}
  \centering
    \includegraphics[width=\columnwidth]{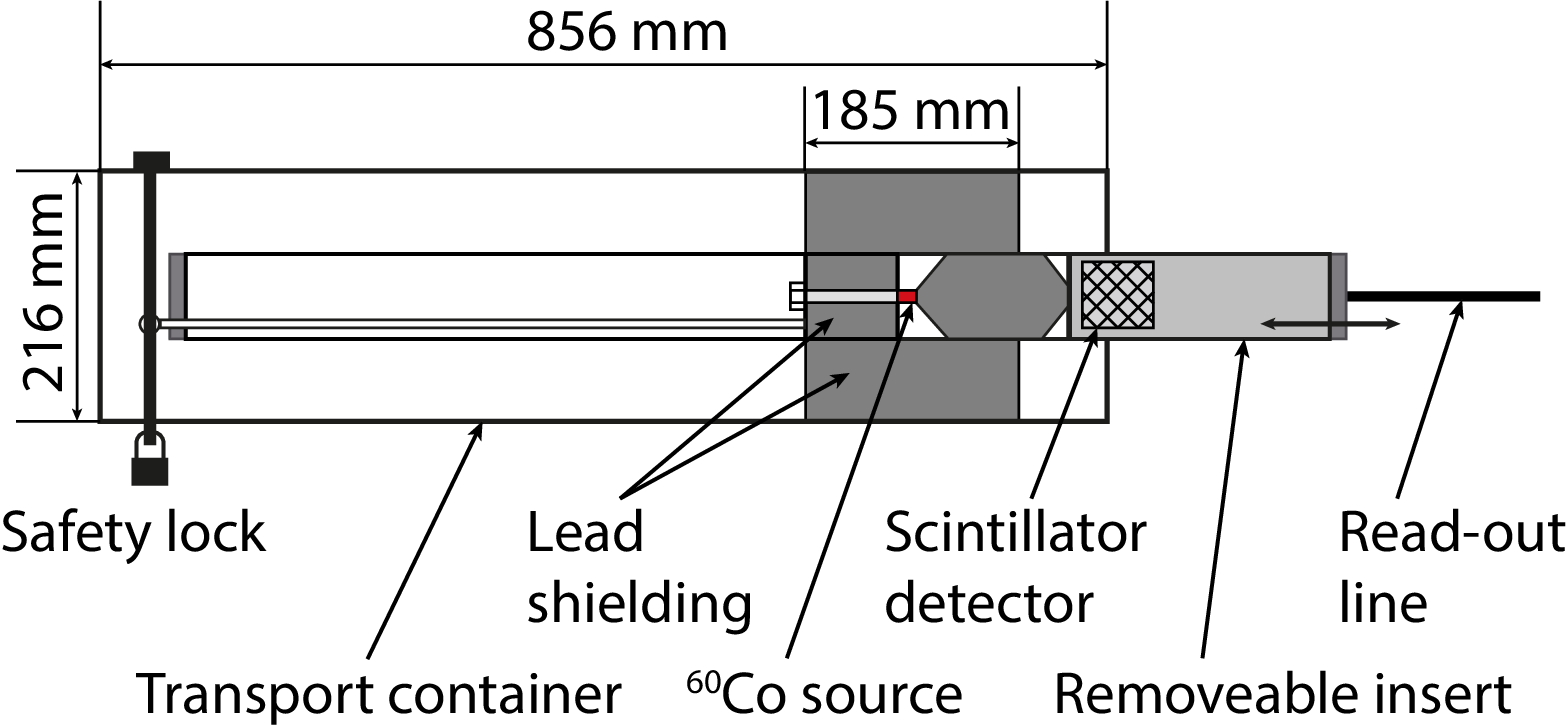}\\
    \vspace{\baselineskip}
    \includegraphics[width=\columnwidth]{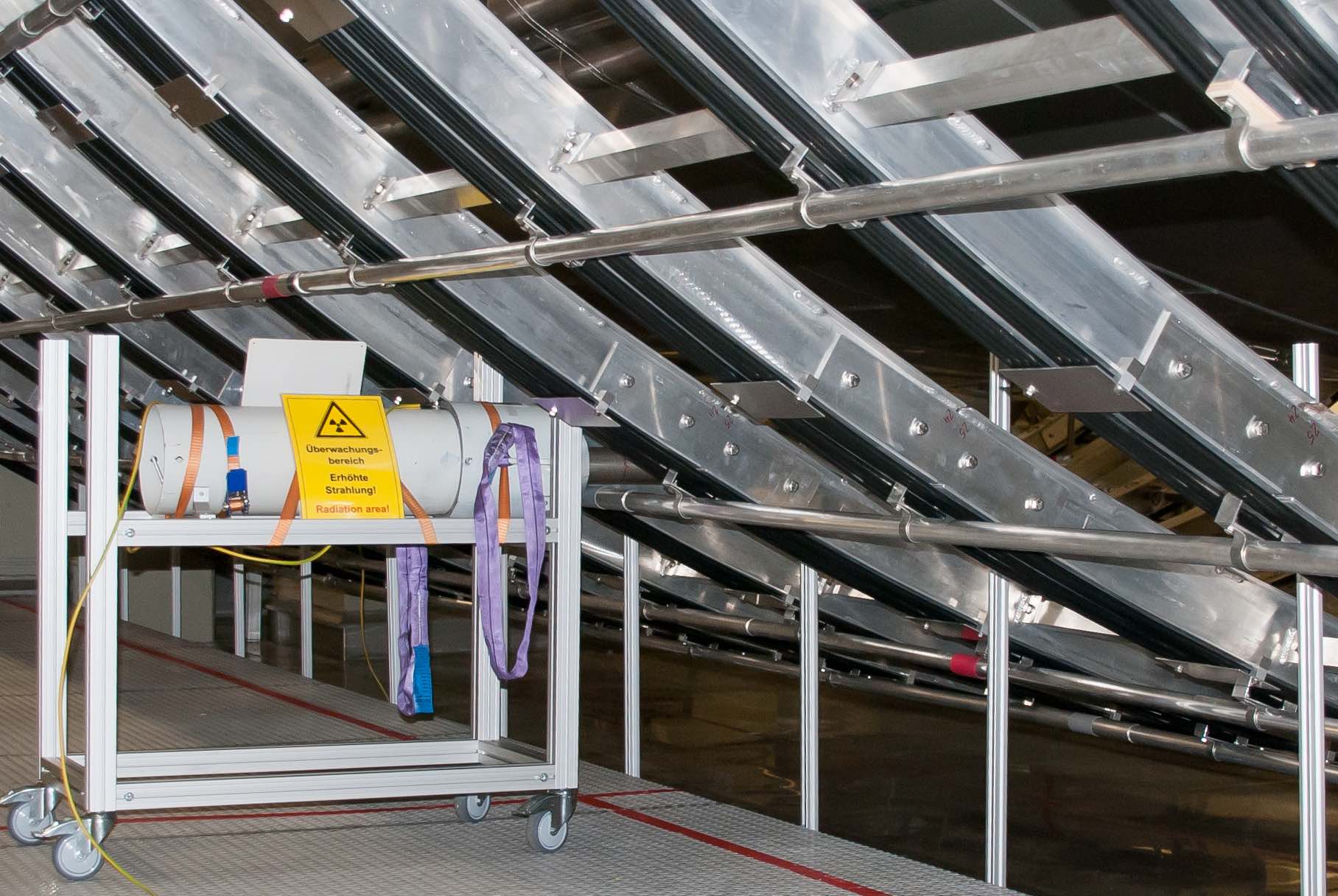}
      \caption{The $^{60}$Co source used to increase the gamma radiation at the MS. 
      The upper panel is a schematic of the source (in the ``closed'' configuration). 
      In the bottom panel, the source can be seen installed next to the air-coils, beneath the MS vessel.}
    \label{fig:CobaltSource}
\end{figure}
     
To increase the gamma-induced background, a $^{60}$Co source with a total activity of \SI[multi-part-units = single]{53.3 \pm 2.7}{\mega\becquerel} was positioned in the vicinity of the MS (Fig.~\ref{fig:CobaltSource}).
The source was originally used for geological surveys along underground piping and is therefore equipped with a shielded scintillator detector. 
It is housed in a lead-shielded transport container that allows convenient handling and transport when not in active use.
Measurements were completed with the source partially outside its lead shielding (``open'' configuration) and completely inside its lead shielding (``closed'' configuration).
%which should reduce the emitted gamma flux by about \num{e3}).
The decay of $^{60}$Co primarily results in the cascade emission of two gammas at \SI{1173}{\kilo\eV} and \SI{1332}{\kilo\eV}~\cite{Browne2013}.  

Measurements of the electron rate with the FPD system were performed with the source located at different locations near the MS.
%Similar results were obtained at the different positions.
However, extended measurements with the ``closed'' configuration were only performed at one position: the $^{60}$Co source located under the west side of the MS, approximately equidistant from the ends of the vessel (see Fig.~\ref{fig:geant4Geometry}).
Therefore, only results from this position are presented here.

\begin{figure*}
    \centering
    \includegraphics[width=\textwidth]{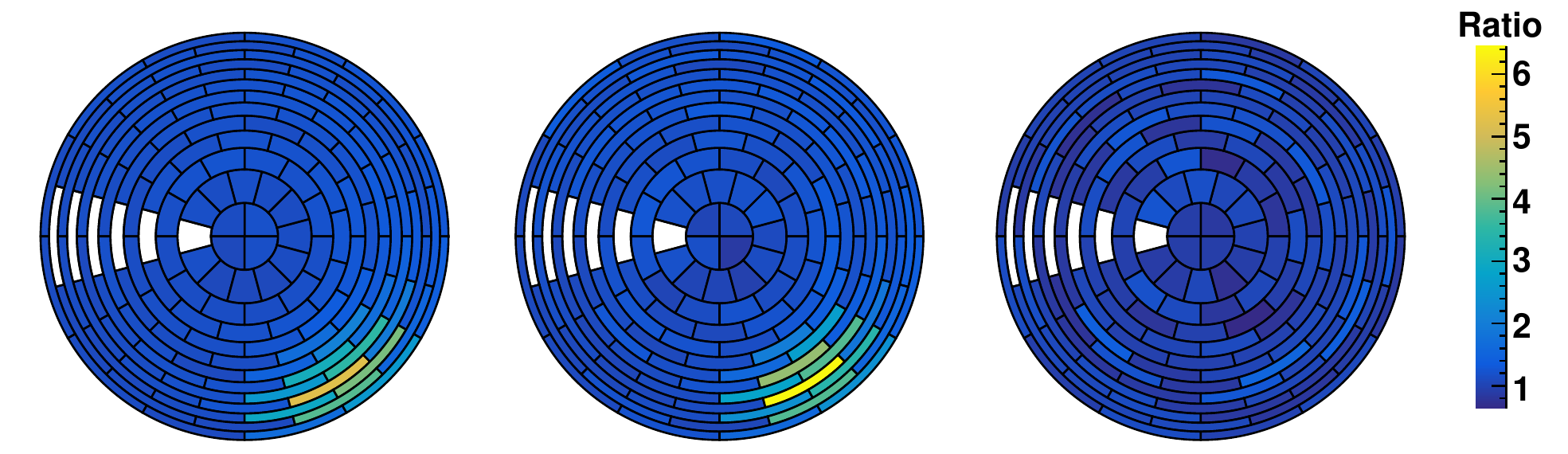}
      \caption{The ratio of the electron rate with the open $^{60}$Co source to the rate with the closed source, for each detector pixel, for three cases: with no magnetic shielding and no electrostatic shielding (left panel), with no magnetic shielding but with electrostatic shielding (center panel), and with both magnetic shielding and electrostatic shielding (right panel).
      The detector is being viewed from the FPD end of the beamline.
      The six white pixels were excluded due to a broken preamplifier module.
      The total FPD rate with the closed source is \SI{683.3}{\cps} (\SI{73.7}{\cps}) (\SI{0.540}{\cps}) for the left (center) (right) panel.
}
    \label{fig:cobaltFPD}
     \end{figure*}

The FPD rate was measured for several magnetic-field and electrostatic shielding configurations; the rates can be found in Table~\ref{table:CobaltRates}.
The effect of the $^{60}$Co source can clearly be seen in the bottom-right portion of the detector wafer in the left and center panels of Fig.~\ref{fig:cobaltFPD} (asymmetric magnetic field), but is absent in the right panel (symmetric magnetic field).

\begin{table*}
      \caption{Results from the FPD measurements with the $^{60}$Co source, where the rates are measured in units of counts per second (cps).
      Rate is the raw result from the detected counts and measurement duration $\Updelta t$, while Rate$^*$ is corrected for the broken preamplifier module.
      The errors on the rates are statistical.
      The magnetic field settings used are those shown in Fig.~\ref{fig:MagneticFieldSettings}.
      For the Asym. M measurements, all rates exclude the inner 16 detector pixels.}
     \label{table:CobaltRates}
    \begin{tabular*}{\textwidth}{@{\extracolsep{\fill}}lccccccccc@{}}
    \hline
     B-field & $U_{\textrm{0}}$ (kV) & $\Updelta U_{\textrm{IE}}$ (V) & Source & Counts & $\Updelta t$ (s) & Rate (\si{\cps}) & $\Updelta$Rate (\si{\cps}) & Rate$^*$ (\si{\cps}) & $\Updelta$Rate$^*$ (\si{\cps}) \\
  \hline
    \multirow{2}{*}{Asym. M} & \multirow{2}{*}{\num{-18.5}} & \multirow{2}{*}{\num{0}} & open & 1500629 & 1790 & \num{838.3 \pm 0.7} & \multirow{2}{*}{\num{222.4 \pm 0.9}} & \num{871.9 \pm 0.7} & \multirow{2}{*}{\num{231.7 \pm 0.9}}\\ 
    & & & closed & 1102527 & 1790 & \num{615.9 \pm 0.6} & & \num{640.2 \pm 0.6} & \\ 
    \hline 
    \multirow{2}{*}{Asym. M} & \multirow{2}{*}{\num{-18.5}} & \multirow{2}{*}{\num{-100}} & open & 164532 & 1790 & \num{91.9 \pm 0.2} & \multirow{2}{*}{\num{25.8 \pm 0.3}} & \num{95.6 \pm 0.2} & \multirow{2}{*}{\num{26.8 \pm 0.3}}\\ 
    & & & closed & 118376 & 1790 & \num{66.1 \pm 0.2} & & \num{68.8 \pm 0.2} & \\ 
    \hline 
    \multirow{2}{*}{Sym.} & \multirow{2}{*}{\num{-18.5}} & \multirow{2}{*}{\num{-100}} & open & 18254 & 33500 & \num{0.545 \pm 0.004} & \multirow{2}{*}{\num{0.005 \pm 0.006}} & \num{0.566 \pm 0.004} & \multirow{2}{*}{\num{0.005 \pm 0.006}}\\  
    & & & closed & 14761 & 27320 & \num{0.540 \pm 0.004} & & \num{0.561 \pm 0.005} & \\ 
    \hline
    \end{tabular*}
\end{table*}

     % Prevents text from going into the margins
     %See https://www.tug.org/tutorials/latex2e/$5csloppy.html
     \sloppy
     
The measurements with the $^{60}$Co source demonstrate the effectiveness of the shielding inside the MS against gamma-induced electron backgrounds.
The background rate due to the $^{60}$Co source under an asymmetric magnetic configuration dropped from \SI[multi-part-units = single]{231.7 \pm 0.9}{\cps} to \SI[multi-part-units = single]{26.8 \pm 0.3}{\cps} with the addition of electrostatic shielding (changing $\Updelta U_{\textrm{IE}}$ from \SI{0}{\V} to \SI{-100}{\V}).
A further reduction in the rate by at least three orders of magnitude occurred when switching to the symmetric magnetic configuration (\SI[multi-part-units = single]{0.005 \pm 0.006}{\cps}, corresponding to less than \SI{1}{\%} of the total background rate).
Though a large number of $^{60}$Co-induced secondary electrons were emitted from the MS surface, no significant rate effect was observed with the nominal magnetic field setting.

\fussy

\begin{table}
      \caption{
      Total gamma fluxes through the interior of the MS as determined from simulations.
      $\Phi_0$ is either the flux with the closed $^{60}$Co source (second column) or with no water shielding (third column).
      Similarly, $\Phi_1$ is either the flux with the open $^{60}$Co source or with water shielding.
      The flux includes all crossings (ingoing and outgoing) of the inner surface.
      The errors include both statistical and systematic uncertainties, although the latter only includes the uncertainty on the activity of the gamma sources (see Table~\ref{table:activities}).
      Other systematic effects, such as the accuracy of the included \textsc{Geant4} physics processes and the correctness of the simulation geometry, were not calculated.
      }
     \label{table:simulationResults}
    \begin{tabular*}{\columnwidth}{@{\extracolsep{\fill}}lcc@{}}
    \hline 
Flux (\SI{e6}{gammas\per\s}) & $^{60}$Co & Water \\ 
\hline
$\Phi_0$ & \num{2.8 \pm 0.1} & \num{2.564 \pm 0.090} \\ 
$\Phi_1$ & \num{20.9 \pm 0.9} & \num{2.411 \pm 0.088}  \\ 
$\Phi_1-\Phi_0$ & \num{18.2 \pm 0.9} & \num{-0.153 \pm 0.002} \\ 
\hline 
    \end{tabular*}
\end{table}

\begin{figure}
	\centering
        \includegraphics[width=\columnwidth]{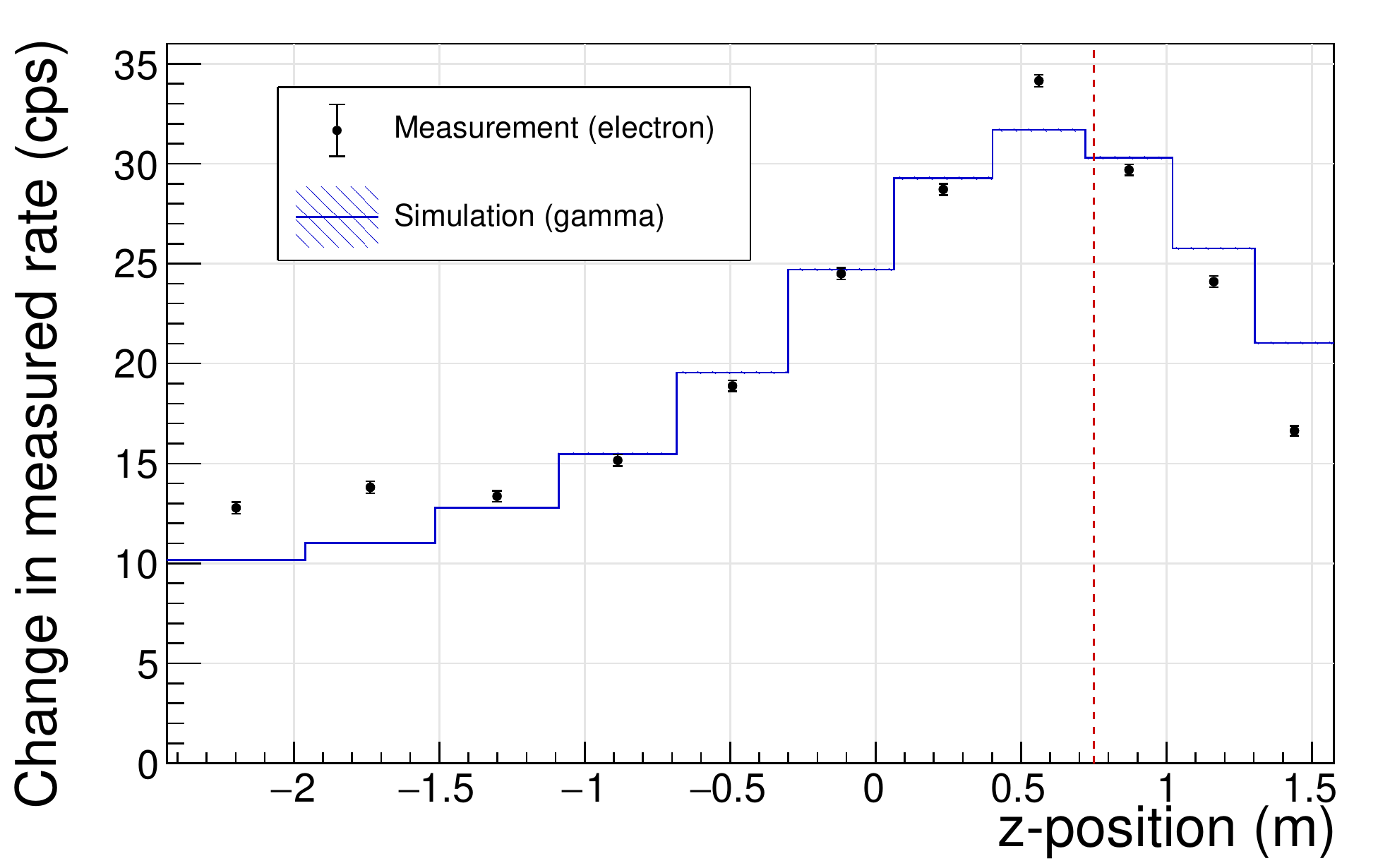}
        \caption{The excess FPD rate measured with the Asym. M magnetic field setting with the $^{60}$Co source, as a function of the axial position $z$.
        Each bin corresponds to a detector pixel ring which images a specific axial region of the MS surface.
        Also shown is the simulated gamma flux induced by the source at the inner surface of the MS, which has been scaled to match the total measured rate.
        The error bars for the measurement and simulation data are statistical only; each of the bin errors for the simulation data is less than \SI{0.4}{\%} of the bin content and so the error bars are difficult to see in the figure.  
        The systematic errors for the simulation data arising from the activity of the $^{60}$Co source are correlated bin-to-bin and are therefore not shown.
        The location of the $^{60}$Co source in the simulation geometry is marked by the dashed line.
	}
        \label{fig:zCobalt}
\end{figure}

A simulation of the gamma flux from the $^{60}$Co source was performed using the geometry described in Section~\ref{sec:simdetails}.
Table~\ref{table:simulationResults} shows the simulated fluxes for gammas traversing the inner surface of the MS vessel.
The presence of the $^{60}$Co source increases the gamma flux through the entire MS surface by about a factor of 8.
The change in flux due to the open $^{60}$Co source is plotted as a function of axial position along the MS in Fig.~\ref{fig:zCobalt}.
The distributions for the measured electron rate (asymmetric field setting) and the simulated gamma flux exhibit similar shapes and peak at the same axial position.

\subsection{Suppression of gamma flux}
\label{subsec:water-suppression}

\begin{figure}
    \centering
    \includegraphics[width=\columnwidth]{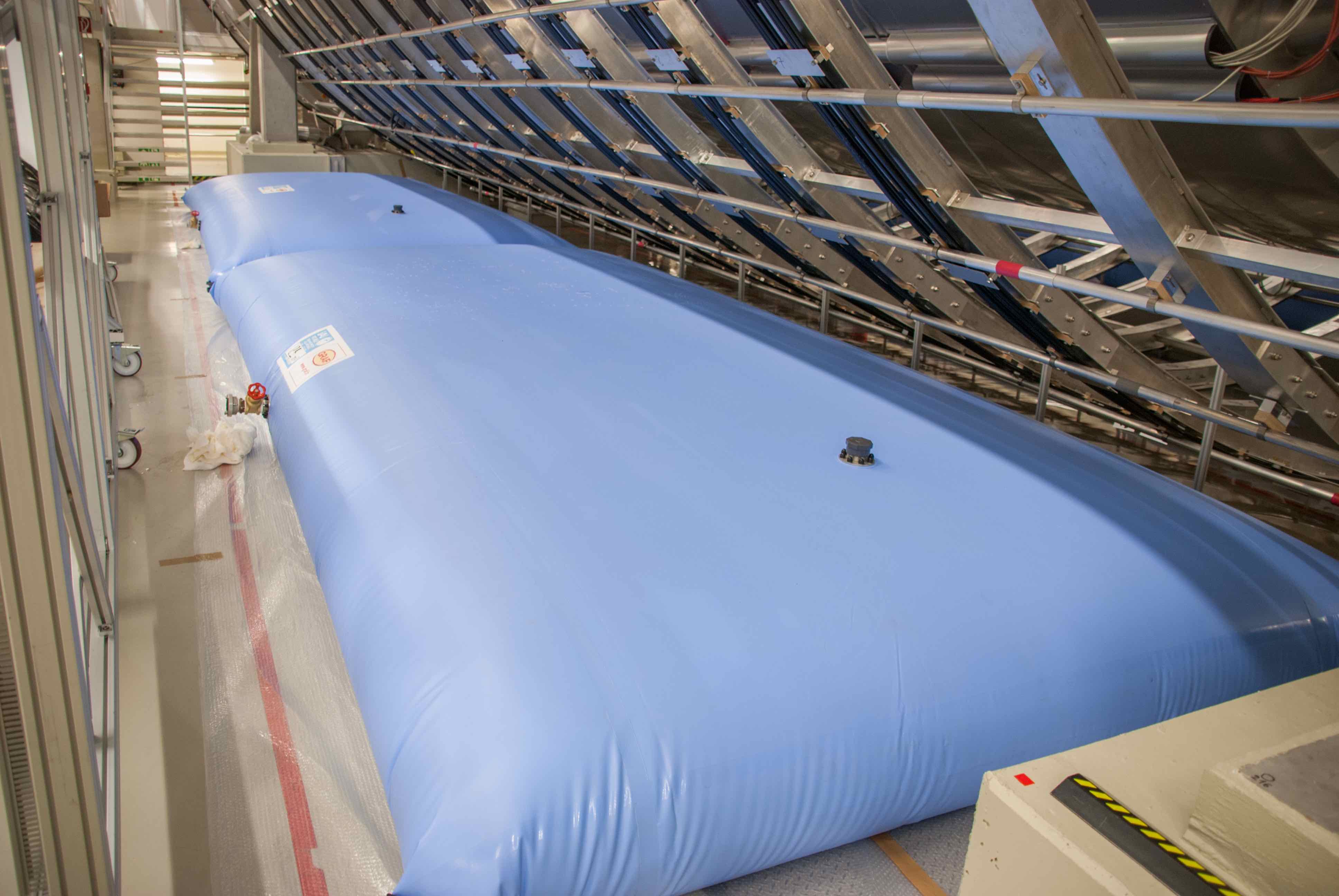}
      \caption{Two of the four flexible water tanks placed beneath the MS during the gamma suppression measurements.}
    \label{fig:WaterShielding}
     \end{figure}

In an effort to reduce the gamma flux originating from the bottom floor of the spectrometer hall, water shielding was temporarily added below the MS.
The basin beneath the MS (\SI{24.1}{m} long by \SI{5.6}{m} wide) was filled with water to a depth of \SI{20}{cm}.
Additionally, a total of four flexible water tanks (each approximately \SI{6.5}{m} long by \SI{3.2}{m} wide by \SI{0.6}{m} high) were installed next to the basin to increase the shielded area.
The water tanks can be seen in Fig.~\ref{fig:WaterShielding}.  

\begin{table*}
 \caption{Results from the FPD measurements with(out) the water shielding. 
      ``Asym. U'' and ``Asym. D'' indicate that the asymmetric magnetic field images the upstream ($z=$\SIrange{-3.9}{-1.0}{\m}) and downstream ($z=$\SIrange{1.3}{3.7}{\m}) region of the MS surface, respectively (see \autoref{fig:MagneticFieldSettings}).
      Rate is the raw result from the detected counts and measurement duration $\Updelta t$, while Rate$^*$ is corrected for the missing detector pixels.
      For the Asym. U (Asym. D) measurements, all rates exclude the inner 28 (4) pixels.
      The errors on the rates are statistical.}
     \label{table:WaterRates}
    \begin{tabular*}{\textwidth}{@{\extracolsep{\fill}}lccccccccc@{}}
     \hline 
     B-field & $U_{\textrm{0}}$ (kV) & $\Updelta U_{\textrm{IE}}$ (V) & Shielding & Counts & $\Updelta t$ (s) & Rate (\si{\cps}) & $\Updelta$Rate (\si{\cps}) & Rate$^*$ (\si{\cps}) & $\Updelta$Rate$^*$ (\si{\cps}) \\
    \hline 
    \multirow{2}{*}{Asym. U} & \multirow{2}{*}{\num{-18.6}} & \multirow{2}{*}{\num{0}} & no water & 21319281 & 35900 & \num{593.9 \pm 0.1} & \multirow{2}{*}{\num{-2.2 \pm 0.2}} & \num{620.1 \pm 0.1} & \multirow{2}{*}{\num{-2.3 \pm 0.2}}\\
    & & & water & 21241447 & 35900 & \num{591.7 \pm 0.1} & & \num{617.8 \pm 0.1} & \\ 
    \hline
    \multirow{2}{*}{Asym. D} & \multirow{2}{*}{\num{-18.6}} & \multirow{2}{*}{\num{0}} & no water & 13767972 & 21990 & \num{626.1 \pm 0.2} & \multirow{2}{*}{\num{-2.6 \pm 0.2}} & \num{653.7 \pm 0.2} & \multirow{2}{*}{\num{-2.8 \pm 0.2}}\\
    & & & water & 24620183 & 39490 & \num{623.5 \pm 0.1} & & \num{650.9 \pm 0.1} & \\  
    \hline
    \multirow{2}{*}{Sym.} & \num{-18.5} & \num{-100} & no water & 19411 & 43080 & \num{0.451 \pm 0.003} & \multirow{2}{*}{\num{0.0008 \pm 0.0043}} & \num{0.469 \pm 0.003} & \multirow{2}{*}{\num{0.0008 \pm 0.0045}}\\ 
    & \num{-18.6} & \num{-100} & water & 25927 & 57440 & \num{0.451 \pm 0.003} & & \num{0.469 \pm 0.003} &\\ 
    \hline
    \end{tabular*}
\end{table*}

The background rates for measurements with and without water shielding are shown in Table~\ref{table:WaterRates}.
Two asymmetric field settings, with field lines intersecting different regions of the MS surface, were implemented; similar reductions in the electron rate due to the water shielding were found for both (\SI{\sim0.4}{\percent}). 
For the symmetric magnetic field setting, the shielding had no significant effect on the electron rate.

\begin{figure}
	\centering
	\includegraphics[width=\columnwidth]{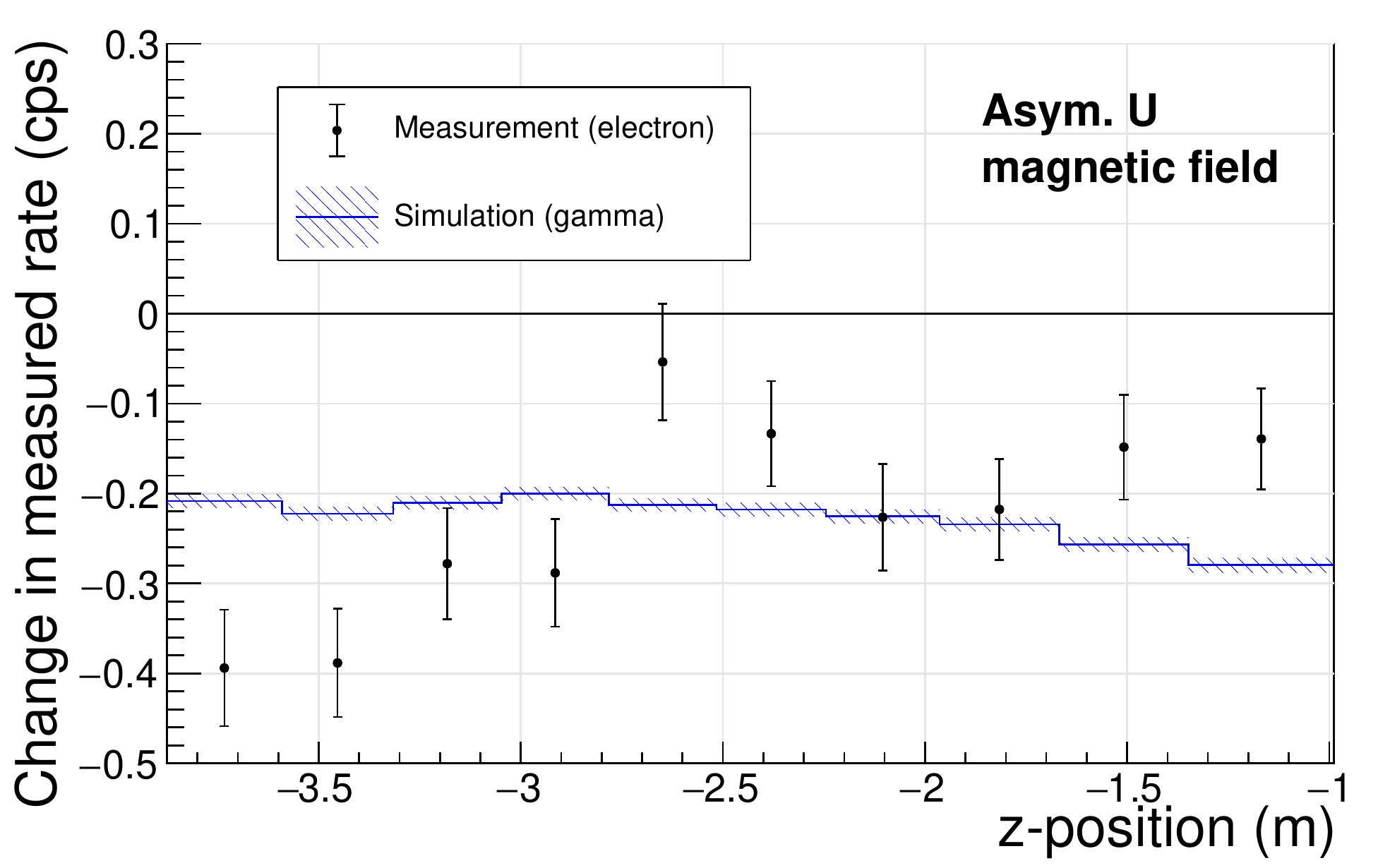}
        \includegraphics[width=\columnwidth]{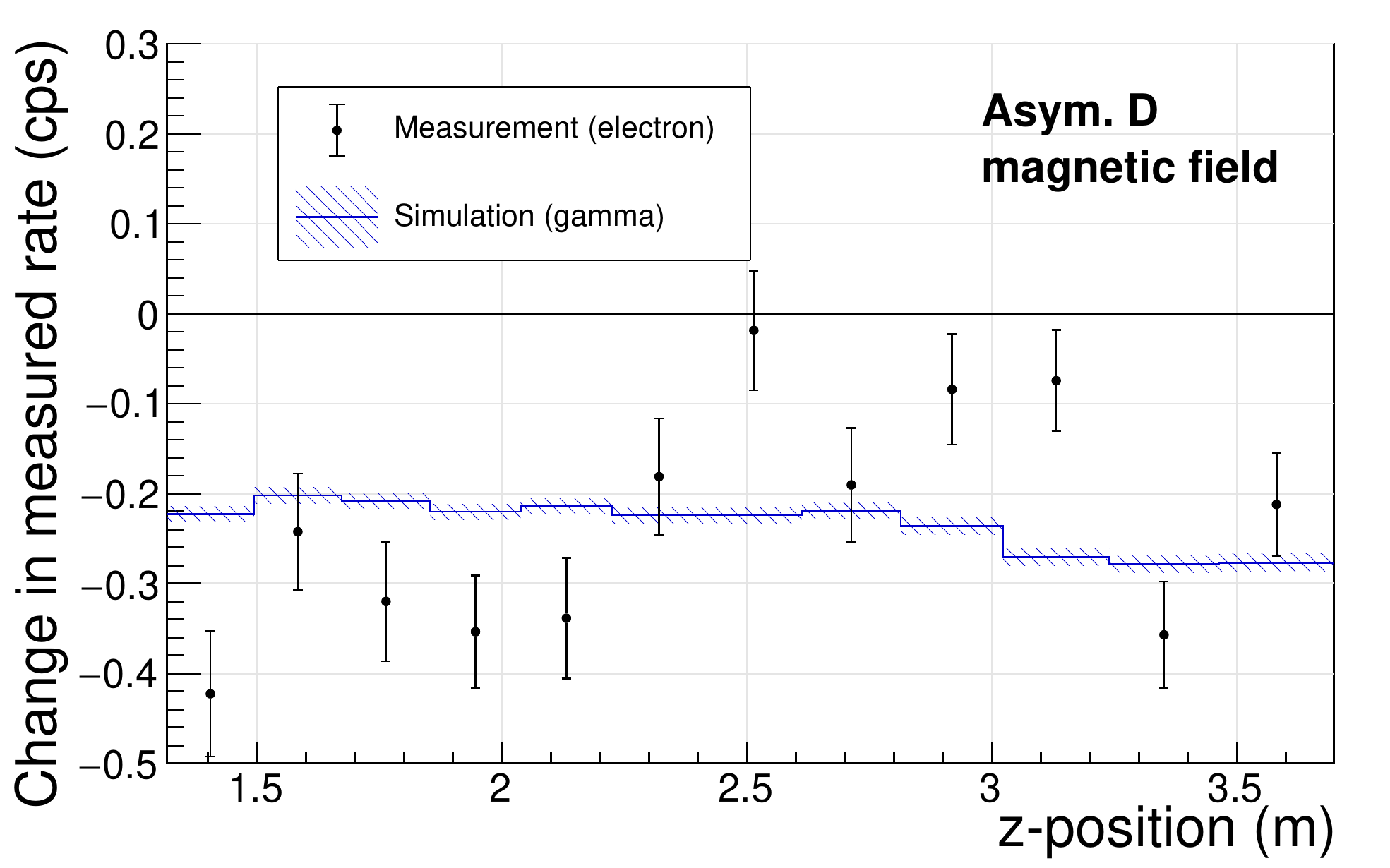}
        \caption{The reduction in the electron rate at the inner surface of the MS, caused by the water shielding, for two asymmetric magnetic field settings.
        The region shielded by the water tanks is roughly $z=$\SIrange{-6.5}{6.5}{\m}.
        The simulated decrease in the gamma flux is also shown in each figure, normalized to the measured rate.
        The error bars for the measurement and simulation data are statistical only.
	}
        \label{fig:zWater}
\end{figure}

To investigate the effect of the shielding on the gamma flux, water was added to the simulation geometry using the dimensions cited above (see Fig.~\ref{fig:geant4Geometry}). 
The simulated fluxes are shown in Table~\ref{table:simulationResults}.  
The addition of water shielding reduced the gamma flux inside the MS by about \SI{7}{\percent}.
The change in flux due to the water shielding is plotted as a function of axial position along the MS in Fig.~\ref{fig:zWater}.
The simulated gamma flux is mostly flat across the measured range and roughly matches the measured electron rate distribution (which is statistics-limited).

\section{Gamma-induced background contribution}
\label{sec:UpperLimit}

By combining the asymmetric magnetic field measurements and simulation results, it is possible to compute two quantities of interest: the secondary electron yield for gamma radiation traversing the inner surface of the MS, and the fraction of secondary electrons caused by environmental gamma radiation.
The symmetric field measurements provide a way to determine the gamma-induced background contribution under standard operating conditions in future $m_{\overline{\nu}_e}$ measurements.

The relevant values used to calculate these quantities are listed in Table~\ref{table:GammaAnalysis}.
Three different rates are to be distinguished: the measured FPD electron rate ($R$), the calculated gamma-induced MS electron emission rate ($S$), and the calculated gamma flux through the MS ($\Phi$).
The values of $R$, $S$, and $\Phi$ can be defined as follows:
\begin{equation}
\begin{aligned}
R &= R_\text{env} + R_\text{other},\\
S &= S_\text{env} + S_\text{other}, \\
\Phi &= \Phi_\text{env} + \Phi_\text{other}\\
\end{aligned}
 \label{eq:FpdRateDefinition}
\end{equation}
where ``env'' indicates the contribution from environmental gamma emitters (e.g. concrete) and ``other'' indicates the contribution from other backgrounds (e.g. cosmic-ray muons).
\autoref{eq:FpdRateDefinition} requires that the $^{60}$Co source is closed and no water shielding is present.

The change in rate due to the $^{60}$Co source is defined to be
\begin{equation}
\Delta R_\text{cobalt} = R_\text{open} - R_\text{closed},
 \label{eq:DeltaFpdRateCobalt}
\end{equation}
while for water shielding the change in rate is
\begin{equation}
\Delta R_\text{shielding} = R_\text{water} - R_\text{no water}.
 \label{eq:DeltaFpdRateWater}
\end{equation}
Identical formulations hold for $\Delta S$ and $\Delta \Phi$. 

%From measurements with the $^{60}$Co source and water shielding, the values of $R$ and $\Delta R$ are easily calculable.
%The gamma simulations provide $\Phi$ and $\Delta \Phi$ values.

\begin{table*}
 \caption{Values used in the calculation of the secondary electron yield ($Y$) and fraction of secondaries induced by environmental gamma radiation ($f_\text{env}$).
 See the text for details about each listed parameter. 
 $\Phi_\text{env}$ and $\Delta\Phi$ differ for each asymmetric field setting (since different regions of the spectrometer surface are measured for each setting), and these values also differ from the values given in \autoref{table:simulationResults} (since the latter considered the entire spectrometer surface).
}
     \label{table:GammaAnalysis}
         \begin{tabular*}{\textwidth}{@{\extracolsep{\fill}}lcccc@{}}
 \hline 
  & $^{60}$Co source & $^{60}$Co source & Water shielding & Water shielding \\
 \hline 
 Magnetic field & Asym. M & Asym. M & Asym. U & Asym. D \\
 z-range (m) & \numrange{-2.44}{1.58} & \numrange{-2.44}{1.58} & \numrange{-3.88}{-0.99} & \numrange{1.32}{3.70} \\
  $\Updelta U_{\textrm{IE}}$ (V) & \num{0} & \num{-100} & \num{0} & \num{0} \\
  \hline
 $R$ (cps) & \num{640.2 \pm 0.6} & \num{68.8 \pm 0.2} & \num{620.1 \pm 0.1} & \num{653.7 \pm 0.2} \\
 $\Delta R$ (cps) & \num{231.7 \pm 0.9} & \num{26.8 \pm 0.3} & \num{-2.3 \pm 0.2} & \num{-2.8 \pm 0.2} \\
  $\Phi_\text{env}$ (\num{e5} gammas/s) & \num{4.5 \pm 0.2} & \num{4.5 \pm 0.2} & \num{3.3 \pm 0.1} & \num{2.7 \pm 0.1} \\
 $\Delta\Phi$ (\num{e5} gammas/s) & \num{86 \pm 4} & \num{86 \pm 4} & \num{-0.257 \pm 0.004} & \num{-0.209 \pm 0.003} \\
%$S$ (e$^-$/s) & \num{0} & \num{0} & \num{0} & \num{0} \\
 %$\Delta S$ (e$^-$/s) & \num{2553 \pm 25} & -- & \num{-19.8 \pm 1.7} & \num{-16.4 \pm 1.3} \\
 %\hline
  $P_\text{arrival}$ (\%) & \num{8.8 \pm 0.1} & -- & \num{11.2 \pm 0.1} & \num{16.6 \pm 0.1} \\
    \hline
 $Y$ (\num{e-4} e$^-$/gamma) & \num{3.1 \pm 0.2} & -- & \num{7.9 \pm 0.7} & \num{8.0 \pm 0.6} \\
  $f_\text{env} (\num{e-2})$ & \num{1.9 \pm 0.1} & \num{2.1 \pm 0.1} & \num{4.6 \pm 0.4} & \num{5.5 \pm 0.5} \\
 \hline 
    \end{tabular*}
\end{table*}

\subsection{Secondary electron yield}
\label{subsec:ProductionRate}

The yield $Y$ is the gamma-induced electron rate divided by the gamma flux through the same surface.
This can be computed from the effect of the $^{60}$Co source or water shielding according to the following equation:
\begin{equation}
Y = \frac{\Delta S}{\Delta\Phi}.
 \label{eq:YieldSource}
\end{equation}
$\Delta S$ can be computed from $\Delta R$ after including electron transport and detection efficiencies:
\begin{equation}
\Delta S = \frac{\Delta R}{\epsilon \cdot P_\text{arrival}},
 \label{eq:EmissionRate}
\end{equation} 
where $\epsilon = \num{0.950 \pm 0.028}$ is the FPD detection efficiency~\cite{Amsbaugh2015} (ignoring the effect of backscattering from the detector surface~\cite{PhDRenschler2011}) and $P_\text{arrival}$ is the average arrival probability for electrons.

Because of the magnetic mirror effect, electrons emitted from the MS surface have a small probability to reach the FPD, which depends on their initial energy and emission angle relative to the magnetic field direction.
$P_\text{arrival}$ was calculated using \textsc{Kassiopeia}~\cite{Furse2017}, the particle-tracking simulation software developed by the KATRIN collaboration.
For each magnetic field configuration, \num{1.6e5} electrons were started on the MS surface with emission angles sampled from a cosine angular distribution~\cite{Henke1977,Seiler1983,Furman2002}.
The electron energy spectrum was assumed to have the following form~\cite{Chung1974, Seiler1983, Joy2004}:
\begin{equation}
 F(E) \propto \frac{E}{(E+W)^{4}},
 \label{eq:Spectrum}
\end{equation}
where $E$ is the electron energy and $W$ = \SI{3.5}{\eV}~\cite{PhDBehrens2016} is the work function of the MS surface.
The energy spectrum for true-secondary electrons is treated independently of the primary particle energy; the shape of the spectrum is known not to vary significantly with respect to the incident photon energy~\cite{Henke1977,Seiler1983}.
%the shape of the spectrum is known to be energy-independent for incident electrons with energies greater than \SI{100}{eV}~\cite{Henke1977,Seiler1983}.
The validity of these assumptions on the electrons' initial properties was shown in the context of the background analysis of cosmic-ray muons~\cite{Altenmueller2019}.

Using the $^{60}$Co measurement, one finds $Y=\SI{3 e-4}{e^{-}/\upgamma}$.
However, the yields derived from the water-shielding measurements give consistent values which are a factor of 2.6 larger than the $^{60}$Co result (see Table~\ref{table:GammaAnalysis}).
One can attempt to explain the discrepancy between the two measurements by recognizing that although the shape of the secondary electron spectrum is independent of the gamma energy, the scaling factor for the electron spectrum is energy dependent~\cite{Grudskii1984}.
Because the gammas emitted from the $^{60}$Co source and the gammas blocked by the water shielding have ostensibly different spectral shapes, an accompanying disparity in electron yields would be expected.
However, an analysis of the simulation results does not substantiate this expectation; inside the spectrometer, the gammas from the $^{60}$Co source and those blocked by the water shielding have very similar average energies (\SI{676}{keV} and \SI{674}{keV}, respectively).

The difference in measured yields is likely the result of an incorrect value for $\Delta\Phi$ obtained from simulations; the tension can be alleviated by decreasing the simulated effect of the $^{60}$Co source or by increasing the simulated effect of the water shielding.  
The latter possibility would entail that the gamma emission from the concrete floor is underestimated in the simulations.
%there are some measurements with the HPGe detector seem to indicate.
%This closeness is energy is due to the fact that most of the high-energy gamma peaks have degraded to low-energy by the time they have passed through the spectrometer steel, and these low-energy gammas dominate
%From simulations, the ratio of the gamma flux emitted from the basement wall to that from the basement floor is \num{3.3}.
%However, the measurements with the HPGe detector (see Section~\ref{sec:siminput_radioactivity}) found a ratio of 1.5.
%These ratios differ by a factor of 2.2, which indicates that the gamma emission from the floor is underestimated in the simulations by a similar factor.
%Therefore, the yield obtained from the $^{60}$Co measurements is the more probable value.

\subsection{Fraction of secondary electrons induced by gammas}
\label{subsec:GammaFraction}

The fraction of secondary electrons emitted from the MS surface which are caused by environmental gamma radiation can be computed in the following manner:
\begin{equation}
f_\text{env} = \frac{S_\text{env}}{S} \approx \frac{R_\text{env}}{R}.
\label{eqn:GammaFractionSimple}
\end{equation}
Assuming that $R_\text{env}$ is proportional to the flux of gammas in the MS, the following relation applies:
\begin{equation}
R_\text{env} \approx \frac{\Delta R}{\Delta\Phi}\Phi_\text{env}.
\label{eqn:GammaRateRatio}
\end{equation}
The gamma-induced fraction can thus be obtained by combining Eqns.~\ref{eqn:GammaFractionSimple} and \ref{eqn:GammaRateRatio}:
\begin{equation}
f_\text{env} \approx \frac{\Phi_\text{env}}{\Delta\Phi}\frac{\Delta R}{R}.
\label{eqn:GammaFraction}
\end{equation}

Table~\ref{table:GammaAnalysis} shows the values of $f_\text{env}$ calculated from the $^{60}$Co measurements under the two electrostatic shielding conditions, as well as from the gamma suppression measurements with water shielding.
The results indicate that less than \SI{6}{\percent} of secondary electrons emitted from the MS surface are induced by environmental gammas.
However, the values from the $^{60}$Co and water shielding measurements differ by a factor of 2.5. 
The scale of this discrepancy is equivalent to the difference in the electron yields between the two types of measurements, as discussed in the previous section.

\subsection{Gamma-induced background rate under standard conditions}
\label{subsec:UpperLimit}

Similar to the asymmetric field measurements, it is possible to use Eqn.~\ref{eqn:GammaRateRatio} to determine the effects of environmental gamma radiation under symmetric field conditions.
Applying the measured and simulated rates listed in Table~\ref{table:UpperLimit}, one finds that $R_\text{env} = \SI[multi-part-units = single]{0.7 \pm 0.9}{\milli\cps}$ (millicount per second), which is consistent with zero.
Assuming the rate is Gaussian, one can follow the unified approach~\cite{Feldman1997} and set an upper limit on the gamma-induced background rate, obtaining $R_\text{env} \leq\SI{2.2}{\milli\cps}$~(\SI{90}{\percent\CL}).

\begin{table*}
 \caption{Background rate $R_\text{env}$ induced by environmental gamma radiation under standard conditions (symmetric magnetic field and $\Updelta U_{\textrm{IE}}=\SI{-100}{\V}$).
 The relevant values used to calculate this rate are also listed; the values of $\Delta \Phi$ and $\Phi_\text{env}$ come from \autoref{table:simulationResults}.
 A corrective factor of 2.6 was applied to the upper limit on $R_\text{env}$ for the $^{60}$Co source.}
     \label{table:UpperLimit}
         \begin{tabular*}{\textwidth}{@{\extracolsep{\fill}}lcc@{}}
 \hline 
 & $^{60}$Co source & Water shielding \\ 
 \hline 
 $\Delta \Phi$ (\SI{e5}{\upgamma/s}) & \num{182 \pm 9} & \num{-1.53 \pm 0.02} \\
 $\Phi_\text{env}$ (\SI{e5}{\upgamma/s}) & \num{25.6 \pm 0.9} & \num{25.6 \pm 0.9} \\
 $\Delta R$ (mcps) & \num{5 \pm 6} & \num{0.8 \pm 4.5} \\
 $R_\text{env}$ (mcps) & \num{0.7 \pm 0.9} & \num{-13 \pm 75} \\
 Upper limit (\SI{90}{\percent\CL}) on $R_\text{env}$ (mcps) & \num{5.6} & \num{110} \\
 \hline 
    \end{tabular*}
\end{table*}

However, one must account for the discrepancy in the results between the $^{60}$Co source and water shielding measurements, as mentioned in the previous sections.
A conservative approach is to allow for the possibility that the simulation overestimates the flux of gammas through the MS from the $^{60}$Co source by a factor of \num{2.6}.
In this case, one finds
\begin{equation}
R_\text{env} \leq\SI{5.6}{\milli\cps}~(\SI{90}{\percent\CL})
\label{eqn:GammaBackgroundLimitConservative}
\end{equation}
Given a nominal rate of \SI{561}{\milli\cps}, this result indicates that less than \SI{\sim1}{\percent} of the MS background rate can be attributed to environmental gamma radiation.

A similar procedure was followed with respect to the water shielding data, giving a limit of $R_\text{env} \leq\SI{110}{\milli\cps}~(\SI{90}{\percent\CL})$\footnote{
Here, a corrective factor of $(2.6)^{-1}$ (which accounts for the possibility that the simulation underestimates the effect of the water shielding) was not applied, in order to obtain a conservative result.}, which is a significantly weaker limit than that obtained from the $^{60}$Co measurements.
The weaker limit obtained with the water shielding configuration is a result of the small simulated value of $\Delta \Phi$ and the large measured uncertainty on $\Delta R$.
Therefore, the measurements with the $^{60}$Co source are more sensitive to the effect of environmental gammas than the water shielding measurements.

\section{Discussion and conclusions}
\label{sec:conclusion}

Low-energy background electrons produced inside the MS are indistinguishable from signal $\upbeta$-particles.
Thus, it is necessary to understand and limit the various sources of electrons in the MS.
Measurements using the asymmetric magnetic field, combined with simulation, indicate that less than \SI{6}{\percent} of secondary electrons emitted from the MS surface are induced by environmental gammas.

The measurements with the $^{60}$Co source show that the electrostatic and magnetic shielding are highly effective in mitigating the effect from secondary electrons.
Changes in the flux of environmental gammas have little if any effect on the MS background rate under standard operating conditions.  
When combined with simulation, the results indicate that less than 
\SI{5.6}{\milli\cps}~(\SI{90}{\percent\CL}) of the MS background rate is gamma-induced, corresponding to less than \SI{\sim1}{\percent} of the total rate.

The remaining (and much more dominant) background rate seems to be caused by the ionization of Rydberg atoms~\cite{PhDTrost2018,Fraenkle2017}.
These atoms are highly excited neutral atoms which can penetrate the electromagnetic shielding implemented in KATRIN.
Furthermore, thermal radiation at room temperature is energetic enough to ionize these atoms and create background electrons.
The production of these Rydberg atoms is thought to primarily originate from the decay of $^{210}$Pb within the MS walls~\cite{PhDHarms2015}.
Plans to mitigate this background are currently being studied.

\begin{acknowledgements}
We acknowledge the support of Helmholtz Association (HGF), Ministry for Education and Research BMBF (5A17PDA, 05A17PM3, 05A17PX3, 05A17VK2, and 05A17WO3), Helmholtz Alliance for Astroparticle Physics (HAP), and Helmholtz Young Investigator Group (VH-NG-1055) in Germany; Ministry of Education, Youth and Sport (\text{CANAM-LM2011019}), cooperation with the JINR Dubna (3+3 grants) 2017--2019 in the Czech Republic; and the Department of Energy through grants DE-FG02-97ER41020, DE-FG02-94ER40818, \text{DE-SC0004036}, DE-FG02-97ER41033, DE-FG02-97ER41041, DE-AC02-05CH11231, \text{DE-SC0011091}, and \text{DE-SC0019304} in the United States.

\end{acknowledgements}

\bibliographystyle{spphys}
\bibliography{GammaBackgroundPaper}

\end{document}

%% file: authors_epjc.tex
% autogenerated by authorlist.py from input file 'overleaf_author_list.tex' using svjour3 format

% Affiliations:
\institute{%
Technische Universit\"{a}t M\"{u}nchen, James-Franck-Str. 1, 85748 Garching, Germany\label{a}
\and IRFU, CEA, Universit\'{e} Paris-Saclay, 91191 Gif-sur-Yvette, France\label{b}
\and Helmholtz-Institut f\"{u}r Strahlen- und Kernphysik, Rheinische Friedrich-Wilhelms Universit\"{a}t Bonn, Nussallee 14-16, 53115 Bonn, Germany\label{c}
\and Karlsruhe Institute of Technology~(KIT), Institute of Experimental Particle Physics~(ETP), Wolfgang-Gaede-Str. 1, 76131 Karlsruhe, Germany\label{d}
\and Institut f\"{u}r Physik, Johannes-Gutenberg-Universit\"{a}t Mainz, 55099 Mainz, Germany\label{e}
\and Karlsruhe Institute of Technology~(KIT), Institute for Data Processing and Electronics~(IPE), Postfach 3640, 76021 Karlsruhe, Germany\label{f}
\and Karlsruhe Institute of Technology~(KIT), Institute for Nuclear Physics~(IKP), Postfach 3640, 76021 Karlsruhe, Germany\label{g}
\and Institut f\"{u}r Kernphysik, Westf\"{a}lische Wilhelms-Universit\"{a}t M\"{u}nster, Wilhelm-Klemm-Str. 9, 48149 M\"{u}nster, Germany\label{h}
\and Institute for Nuclear Research of Russian Academy of Sciences, 60th October Anniversary Prospect 7a, 117312 Moscow, Russia\label{i}
\and Karlsruhe Institute of Technology~(KIT), Institute for Technical Physics~(ITeP), Postfach 3640, 76021 Karlsruhe, Germany\label{j}
\and Max-Planck-Institut f\"{u}r Kernphysik, Saupfercheckweg 1, 69117 Heidelberg, Germany\label{k}
\and Max-Planck-Institut f\"{u}r Physik, F\"{o}hringer Ring 6, 80805 M\"{u}nchen, Germany\label{l}
\and Laboratory for Nuclear Science, Massachusetts Institute of Technology, 77 Massachusetts Ave, Cambridge, MA 02139, USA\label{m}
\and Center for Experimental Nuclear Physics and Astrophysics, and Dept.~of Physics, University of Washington, Seattle, WA 98195, USA\label{n}
\and Nuclear Physics Institute of the CAS, v.~v.~i., CZ-250 68 \v{R}e\v{z}, Czech Republic\label{o}
\and Department of Physics, Faculty of Mathematics und Natural Sciences, University of Wuppertal, Gauss-Str. 20, 42119 Wuppertal, Germany\label{p}
\and Department of Physics, Carnegie Mellon University, Pittsburgh, PA 15213, USA\label{q}
\and Universidad Complutense de Madrid, Instituto Pluridisciplinar, Paseo Juan XXIII, n\textsuperscript{\b{o}} 1, 28040 - Madrid, Spain\label{r}
\and Department of Physics and Astronomy, University of North Carolina, Chapel Hill, NC 27599, USA\label{s}
\and Triangle Universities Nuclear Laboratory, Durham, NC 27708, USA\label{t}
\and Institute for Nuclear and Particle Astrophysics and Nuclear Science Division, Lawrence Berkeley National Laboratory, Berkeley, CA 94720, USA\label{u}
\and University of Applied Sciences~(HFD)~Fulda, Leipziger Str.~123, 36037 Fulda, Germany\label{v}
\and Department of Physics, Case Western Reserve University, Cleveland, OH 44106, USA\label{w}
\and Institut f\"{u}r Physik, Humboldt-Universit\"{a}t zu Berlin, Newtonstr. 15, 12489 Berlin, Germany\label{x}
\and Karlsruhe Institute of Technology~(KIT), Project, Process, and Quality Management~(PPQ), Postfach 3640, 76021 Karlsruhe, Germany\label{y}
}

% Authors:
\author{%
K.~Altenm\"{u}ller\thanksref{a,b}
\and M.~Arenz\thanksref{c}
\and W.-J.~Baek\thanksref{d}
\and M.~Beck\thanksref{e}
\and A.~Beglarian\thanksref{f}
\and J.~Behrens\thanksref{g,d,h}
\and A.~Berlev\thanksref{i}
\and U.~Besserer\thanksref{j}
\and K.~Blaum\thanksref{k}
\and F.~Block\thanksref{d}
% ...
}

\thankstext{corr}{\email{lkippenb@uw.edu}}
\thankstext{also1}{Also affiliated with Oak Ridge National Laboratory, Oak Ridge, TN 37831, USA}